\documentclass[a4paper,12pt,reqno]{amsart}

\usepackage{graphicx,amssymb,datetime,float,MnSymbol,currfile,tikz,multirow}
\usepackage[square,comma,numbers]{natbib}
\usepackage[foot]{amsaddr}
\usetikzlibrary{arrows}
\usetikzlibrary{decorations.markings}

\def\ci{\!\perp\!}

\def\ra{\rightarrow}

\newcommand{\comments}[1]{}

\tikzset{tt/.style={decoration={
  markings,
  mark=at position .485 with {\arrow{>}},
  mark=at position .515 with {\arrow{<}}},postaction={decorate}}}

\begin{document}

\title[]{On the Probability of Immunity}

\author{Jose M. Pe\~{n}a$^1$}
\address{$^1$Link\"oping University, Sweden.}
\email{jose.m.pena@liu.se}


\maketitle

\begin{abstract}
This work is devoted to the study of the probability of immunity, i.e. the effect occurs whether exposed or not. We derive necessary and sufficient conditions for non-immunity and $\epsilon$-bounded immunity, i.e. the probability of immunity is zero and $\epsilon$-bounded, respectively. The former allows us to estimate the probability of benefit (i.e., the effect occurs if and only if exposed) from a randomized controlled trial, and the latter allows us to produce bounds of the probability of benefit that are tighter than the existing ones. We also introduce the concept of indirect immunity (i.e., through a mediator) and repeat our previous analysis for it. Finally, we propose a method for sensitivity analysis of the probability of immunity under unmeasured confounding.
\end{abstract}

\section{Introduction}

Let $X$ and $Y$ denote an exposure and its outcome, respectively. Let $X$ and $Y$ be binary taking values in $\{x,x'\}$ and $\{y,y'\}$. Let $Y_x$ and $Y_{x'}$ denote the counterfactual outcome when the exposure is set to level $X=x$ and $X=x'$. Let $y_x$, $y'_x$, $y_{x'}$ and $y'_{x'}$ denote the events $Y_x=y$, $Y_x=y'$, $Y_{x'}=y$ and $Y_{x'}=y'$. For instance, let $X$ represent whether a patient gets treated or not for a deadly disease, and $Y$ represent whether she survives it or not. Individual patients can be classified into immune (they survive whether they are treated or not, i.e. $y_x \land y_{x'}$), doomed (they die whether they are treated or not, i.e. $y'_x \land y'_{x'}$), benefited (they survive if and only if treated, i.e. $y_x \land y'_{x'}$), and harmed (they die if and only if treated, i.e. $y'_x \land y_{x'}$).

In general, the average treatment effect (ATE) estimated from a randomized controlled trial (RCT) does not inform about the probability of benefit (or of any of the other response types, i.e. harm, immunity, and doom). However, it may do it under certain conditions. For instance,
\begin{align}\label{eq:MandP}\nonumber
ATE = p(y_x) - p(y_{x'}) &= p(y_x, y_{x'}) + p(y_x, y'_{x'}) - [ p(y_x, y_{x'}) + p(y'_x, y_{x'}) ]\\
&= p(\text{benefit}) - p(\text{harm})
\end{align}
and thus $p(\text{benefit})=ATE$ if $p(\text{harm})=0$ (a.k.a. monotonicity). Necessary and sufficient conditions are derived by \citet{MuellerandPearl2023} to determine from observational and experimental data if monotonicity holds. In this work, we derive similar conditions for non-immunity, i.e. $p(\text{immunity})=p(y_x, y_{x'})=0$. These are interesting because under non-monotonicity, they turn an RCT informative about the probabilities of benefit and harm. To see it, consider
\[
ATE = p(y_x) - p(y_{x'})
\]
where the terms on the right-hand side of the equation are estimated from an RCT. Moreover,
\begin{align}\label{eq:pyx}
p(y_x)&=p(y_x,y_{x'})+p(y_x,y'_{x'})=p(\text{immunity}) + p(\text{benefit})\\\label{eq:pyxp}
p(y_{x'})&=p(y_x,y_{x'})+p(y'_x,y_{x'})=p(\text{immunity}) + p(\text{harm})
\end{align}
and thus $p(\text{benefit})=p(y_x)$ and $p(\text{harm})=p(y_{x'})$ if $p(\text{immunity})=0$.

In some cases, non-immunity is assured. For instance, when evaluating the effect of advertising on the purchase of a new product. The control group not being exposed to the ad has no way of purchasing the product, i.e. $p(y_{x'})=0$ and thus $p(y_x, y_{x'})=0$. In other cases, non-immunity cannot be assured. For instance, when evaluating the effect of a drug. An individual may carry a gene variant that makes her recover from the disease regardless of whether she takes the drug or not, i.e. $p(y_x, y_{x'}) \geq 0$. However, it may still be bounded as $p(y_x, y_{x'}) \leq \epsilon$ from expert knowledge. We show that our necessary and sufficient conditions for non-immunity can trivially be adapted to $\epsilon$-bounded immunity. Moreover, we show that the knowledge of $\epsilon$-bounded immunity may tighten the bounds of the probabilities of benefit and harm by \citet{TianandPearl2000}. We also introduce the concepts of indirect benefit and harm (i.e., through a mediator) and repeat our previous analysis for them. Finally, we propose a method for sensitivity analysis of immunity under unmeasured confounding. We illustrate our results with concrete examples.

\section{Conditions for Non-Immunity}

Consider the bounds of $p(\text{benefit})$ derived by \citet{TianandPearl2000}:
\begin{equation}\label{eq:TianandPearl}
\max \left\{
\begin{array}{cc}
0,\\
p(y_x)-p(y_{x'}),\\
p(y)-p(y_{x'}),\\
p(y_x)-p(y)
\end{array}
\right\}
\leq p(\text{benefit}) \leq
\min \left\{
\begin{array}{cc}
p(y_x),\\
p(y'_{x'}),\\
p(x,y)+p(x',y'),\\
p(y_x)-p(y_{x'})+\\p(x,y')+p(x',y)
\end{array}
\right\}.
\end{equation}
Then, combining Equations \ref{eq:pyx} or \ref{eq:pyxp} with \ref{eq:TianandPearl} gives
\begin{equation}\label{eq:pi}
\max \left\{
\begin{array}{cc}
0,\\
p(y_x)-p(y'_{x'}),\\
p(y_x)-p(x,y)-\\
p(x',y'),\\
p(y_{x'})-p(x,y')-\\
p(x',y)
\end{array}
\right\}
\leq p(\text{immunity}) \leq
\min \left\{
\begin{array}{cc}
p(y_x),\\
p(y_{x'}),\\
p(y_x)-p(y)+\\
p(y_{x'}),\\
p(y)
\end{array}
\right\}.
\end{equation}

A sufficient condition for $p(\text{immunity})=0$ to hold is that some argument to the min function in Equation \ref{eq:pi} is equal to 0, that is
\begin{equation}\label{eq:sufficient}
p(y_x)=0 \text{ or } p(y_{x'})=0 \text{ or } p(y_x) + p(y_{x'}) = p(y) \text{ or } p(y)=0.
\end{equation}
Likewise, a necessary condition for $p(\text{immunity})=0$ to hold is that all the arguments to the max function are non-positive, that is
\begin{align}\label{eq:necessary}\nonumber
&p(y_x) + p(y_{x'}) \leq 1 \text{ and }\\\nonumber
&p(y_x) \leq p(x,y)+p(x',y') \text{ and }\\
&p(y_{x'}) \leq p(x,y')+p(x',y).
\end{align}

\subsection{Conditions for $\epsilon$-Bounded Immunity}\label{sec:cebi}

The conditions in the previous section can be relaxed to allow certain degree of immunity (e.g., based on expert knowledge), making them more applicable in practice. Specifically, a sufficient condition for $p(\text{immunity}) \leq \epsilon$ to hold is
\[
p(y_x) \leq \epsilon \text{ or } p(y_{x'}) \leq \epsilon \text{ or } p(y_x) + p(y_{x'}) \leq p(y) + \epsilon \text{ or } p(y) \leq \epsilon.
\]
Likewise, a necessary condition for $p(\text{immunity}) \leq \epsilon$ to hold is
\begin{align}\label{eq:necessarye}\nonumber
&p(y_x) + p(y_{x'}) \leq 1 + \epsilon \text{ and }\\\nonumber
&p(y_x) \leq p(x,y)+p(x',y') + \epsilon \text{ and }\\
&p(y_{x'}) \leq p(x,y')+p(x',y) + \epsilon.
\end{align}

\subsection{$\epsilon$-Bounds on Benefit and Harm}\label{sec:ebbh}

Assuming $\epsilon$-bounded immunity (e.g., based on expert knowledge) can help narrowing the bounds on $p(\text{benefit})$ and $p(\text{harm})$. Specifically, if $p(\text{immunity}) \leq \epsilon$ then Equation \ref{eq:pyx} gives
\[
p(y_x) - \epsilon \leq p(\text{benefit}) \leq p(y_x).
\]
Incorporating this into Equation \ref{eq:TianandPearl} gives
\begin{equation}\label{eq:pbe}
\max \left\{
\begin{array}{cc}
0,\\
p(y_x)-p(y_{x'}),\\
p(y)-p(y_{x'}),\\
p(y_x)-p(y),\\
p(y_x) - \epsilon
\end{array}
\right\}
\leq p(\text{benefit}) \leq
\min \left\{
\begin{array}{cc}
p(y_x),\\
p(y'_{x'}),\\
p(x,y)+p(x',y'),\\
p(y_x)-p(y_{x'})+\\p(x,y')+p(x',y)
\end{array}
\right\}
\end{equation}
which can potentially return a tighter lower bound than Equation \ref{eq:TianandPearl}, i.e. if $\epsilon < \min(p(y_{x'}), p(y))$. Although the value of $\epsilon$ is typically determined from expert knowledge and not from data, the experimental and observational data available do restrict the values that are valid, as indicated by Equation \ref{eq:necessarye}. In short, $\epsilon$ can take any value as long as the lower bound is not greater than the upper bound in Equation \ref{eq:pbe}. Moreover, $p(\text{harm})$ can likewise be bounded by simply swapping $x$ and $x'$ in Equation \ref{eq:pbe}.

\subsection{Examples}

This section illustrates the results above with two concrete examples.\footnote{\texttt{R} code for the examples can be found at \texttt{https://tinyurl.com/2s3bxmyu}.}

\subsubsection{Example 1}

A pharmaceutical company wants to market their drug to cure a disease by claiming that no one is immune. The RCT they conducted for the drug approval yielded the following:
\begin{align*}
p(y_x) &= 0.76\\
p(y_{x'}) &= 0.31
\end{align*}
which correspond to the following unknown data generation model:
\begin{align*}
p(u)=0.3 && p(x|u)=0.2 && p(y|x,u)&=0.9\\
&&&& p(y|x,u')&=0.7\\
&& p(x|u')=0.9 && p(y|x',u)&=0.8\\
&&&& p(y|x',u')&=0.1.
\end{align*}
Therefore, the necessary condition for non-immunity in Equation \ref{eq:necessary} does not hold, and thus the company is not entitled to make the claim they intended to make. The company changes strategy and now wishes to market their drug as having a minimum of 50 \% efficacy, i.e. benefit. To do so, they first conduct an observational study that yields the following:
\begin{align*}
p(x,y)=0.5 && p(x,y')&=0.2\\
p(x',y)=0.2 && p(x',y')&=0.1.
\end{align*}
Then, they apply Equation \ref{eq:TianandPearl} to the RCT and observational results to conclude that $0.45 \leq p(\text{benefit}) \leq 0.61$. Again, the company cannot proceed with their marketing strategy. A few months later, a research publication reports that no more than 25 \% of the population is immune. The company realizes that this value is compatible with their RCT and observational results, by checking the necessary condition for $\epsilon$-bounded immunity in Equation \ref{eq:necessarye}. More importantly, the company realizes that Equation \ref{eq:pbe} with $\epsilon=0.25$ allows to conclude that $0.51 \leq p(\text{benefit}) \leq 0.61$, and thus they can resume their marketing strategy.

\subsubsection{Example 2}

The previous example has shown that expert knowledge on immunity may complement experimental and observational data. While data alone rarely provide precise information on immunity (or on any other response type, for that matter), there are cases where data alone provide enough actionable information. The following example illustrates this.

A pharmaceutical company is concerned by the poor sales of a drug to cure a disease. The RCT conducted for the drug approval and a subsequent observational study yielded the following:
\begin{align*}
p(y_x) &= 0.48\\
p(y_{x'}) &= 0.36
\end{align*}
and
\begin{align*}
p(x,y)=0.08 && p(x,y')&=0.2\\
p(x',y)=0.25 && p(x',y')&=0.47.
\end{align*}
which correspond to the following unknown data generation model:
\begin{align*}
p(u)=0.4 && p(x|u)=0.1 && p(y|x,u)&=0.9\\
&&&& p(y|x,u')&=0.2\\
&& p(x|u')=0.4 && p(y|x',u)&=0.3\\
&&&& p(y|x',u')&=0.4.
\end{align*}

The fact that 36 \% of the untreated recover from the disease makes the company suspect that the low sales are due to a large part of the population being immune. Equation \ref{eq:pi} allows to conclude that $0 \leq p(\text{immunity}) \leq 0.34$, which suggests that the explanation offered by the company is rather unlikely. A more plausible explanation for the low sales may be that the efficacy or benefit of the drug is not very high, as $0.14 \leq p(\text{benefit}) \leq 0.48$ by Equation \ref{eq:TianandPearl}.

\section{Indirect Benefit and Harm}

In the previous sections, the causal graph of the domain under study was unknown. In this section, we assume that the graph is available (e.g., from expert knowledge) and discuss two advantages that follow with it. Specifically, suppose that the domain under study corresponds to the following causal graph:
\begin{center}
\begin{tikzpicture}[inner sep=1mm]
\node at (0,0) (E) {$X$};
\node at (2,0) (D) {$Y$};
\node at (1,0) (Z) {$Z$};
\path[->] (E) edge (Z);
\path[->] (Z) edge (D);
\path[->] (E) edge[bend left] (D);
\end{tikzpicture}
\end{center}
and thus $p(y_x)=p(y|x)$ and $p(y_{x'})=p(y|x')$. Then, $p(y_x)$ and $p(y_{x'})$ can be estimated from observational data and thus, unlike in the previous sections, no RCT is required. A further advantage is that we can now compute the probabilities of benefit and harm mediated by $Z$. We elaborate on this below.

The effect of $X$ on $Y$ mediated by $Z$ (a.k.a. indirect effect) corresponds to the effect due to the indirect path $X \ra Z \ra Y$, i.e. after deactivating the direct path $X \ra Y$. Different ways of deactivating the direct path have resulted in different indirect effect measures in the literature. \citet{Pearl2001} proposes deactivating the direct path by setting $X$ to non-exposure and comparing the expected outcome when $Z$ takes the value it would under exposure and non-exposure:
\[
NIE=E[Y_{x',Z_x}]-E[Y_{x'}]
\]
which is known as the average natural (or pure) indirect effect. \citet{Geneletti2007} also proposes deactivating the direct path by setting $X$ to non-exposure but instead, she proposes comparing the expected outcome when $Z$ is drawn from the distributions $\mathcal{Z}_x$ and $\mathcal{Z}_{x'}$ of $Z_x$ and $Z_{x'}$:
\[
IIE=E[Y_{x',\mathcal{Z}_x}]-E[Y_{x',\mathcal{Z}_{x'}}]
\]
which is known as the interventional indirect effect. Although $NIE$ and $IIE$ do not coincide in general, they coincide for the causal graph above \cite{VanderWeeleetal.2014}. Finally, \citet{Fulcheretal.2020} proposes deactivating the direct path by setting $X$ to its natural (observed) value and comparing the expected outcome when $Z$ takes its natural value and the value it would under no exposure:
\[
PIIE = E[Y_{X,Z_X}] - E[Y_{X,Z_{x'}}]
\]
which is also known as the population intervention indirect effect. This measure is suitable when the exposure is harmful (e.g., smoking), and thus one may be more interested in elucidating the effect (e.g., disease prevalence) of eliminating the exposure rather than in contrasting the effects of exposure and non-exposure.

We propose an alternative way of deactivating the direct path $X \ra Y$ and measuring the indirect effect of $X$ on $Y$ through $Z$. Specifically, we assume that the direct path $X \ra Y$ is actually mediated by an unmeasured random variable $U$ that is left unmodelled. This arguably holds in most domains. The identity of $U$ is irrelevant. Let $G$ denote the causal graph below, i.e. the original causal graph refined with the addition of $U$.
\begin{center}
\begin{tikzpicture}[inner sep=1mm]
\node at (0,0) (E) {$X$};
\node at (2,0) (D) {$Y$};
\node at (1,1) (U) {$U$};
\node at (1,0) (Z) {$Z$};
\path[->] (E) edge (Z);
\path[->] (Z) edge (D);
\path[->] (E) edge (U);
\path[->] (U) edge (D);
\end{tikzpicture}
\end{center}
Now, deactivating the direct path $X \ra Y$ in the original causal graph can be achieved by adjusting for $U$ in $G$, i.e. $\sum_u E[Y|x,u] p(u)$. Unfortunately, $U$ is unmeasured. Instead, we propose the following way of deactivating $X \ra Y$. Let $H$ denote the causal graph below, i.e. the result of reversing the edge $X \ra U$ in $G$.
\begin{center}
\begin{tikzpicture}[inner sep=1mm]
\node at (0,0) (E) {$X$};
\node at (2,0) (D) {$Y$};
\node at (1,1) (U) {$U$};
\node at (1,0) (Z) {$Z$};
\path[->] (E) edge (Z);
\path[->] (Z) edge (D);
\path[->] (U) edge (E);
\path[->] (U) edge (D);
\end{tikzpicture}
\end{center}
The average total effect of $X$ on $Y$ in $H$ can be computed by the front-door criterion \cite{Pearl2009}:
\begin{align*}
TE &= E[Y_x] - E[Y_{x'}]\\
&= \sum_z p(z|x) \sum_{\dot{x}} E[Y|\dot{x},z] p(\dot{x}) - \sum_z p(z|x') \sum_{\dot{x}} E[Y|\dot{x},z] p(\dot{x}).
\end{align*}
Note that $G$ and $H$ are distribution equivalent, i.e. every probability distribution that is representable by $G$ is representable by $H$ and vice versa \cite{Pearl2009}. Then, evaluating the second line of the equation above in $G$ or $H$ gives the same result. If we evaluate it in $H$, then it corresponds to the part of association between $X$ and $Y$ that is attributable to the path $X \ra Z \ra Y$. If we evaluate it in $G$, then it corresponds to the part of $TE$ in $G$ that is attributable to the path $X \ra Z \ra Y$, because $TE$ in $G$ equals the association between $X$ and $Y$, since $G$ has only directed paths from $X$ to $Y$. Therefore, the second line in the equation above corresponds to the part of $TE$ in the original causal graph that is attributable to the path $X \ra Z \ra Y$, thereby deactivating the direct path $X \ra Y$. We propose to use the second line in the equation above as a measure of the indirect effect of $X$ on $Y$ in the original causal graph.

The reasoning above can be extended to the probabilities of benefit and harm, and thereby measure the benefit and harm mediated by $Z$. As mentioned above, the causal graphs $G$ and $H$ represent different data generation mechanisms but the same probability distribution over $X$, $Y$ and $Z$. Therefore, the mechanisms agree on observational probabilities but may disagree on counterfactual probabilities. We use $p()$ to denote observational probabilities obtained from either mechanism, and $q()$ to denote counterfactual probabilities obtained from the mechanism corresponding to $H$. The probabilities of benefit and harm of $X$ on $Y$ mediated by $Z$ in $G$ and thus in the original causal graph (henceforth indirect benefit and harm, or $IB$ and $IH$) can be computed by applying Equation \ref{eq:pyx} to $H$. That is,
\[
IB=q(\text{benefit})=q(y_x)=\sum_z p(z|x) \sum_{\dot{x}} p(y|\dot{x},z) p(\dot{x})
\]
where the second equality holds if $q(\text{immunity})=0$, and the third is due to the front-door criterion on $H$. Likewise for $IH$ simply replacing $x$ by $x'$. Applying Equation \ref{eq:pi} to $H$ yields necessary and sufficient conditions for $q(\text{immunity})=0$. That is,
\begin{align}\label{eq:sufficienti}\nonumber
&\sum_z p(z|x) \sum_{\dot{x}} p(y|\dot{x},z) p(\dot{x})=0 \text{ or }\\\nonumber
&\sum_z p(z|x') \sum_{\dot{x}} p(y|\dot{x},z) p(\dot{x})=0 \text{ or }\\\nonumber
&\sum_z [ p(z|x) + p(z|x') ] \sum_{\dot{x}} p(y|\dot{x},z) p(\dot{x}) = p(y) \text{ or }\\
&p(y)=0
\end{align}
is a sufficient condition, whereas 
\begin{align}\label{eq:necessaryi}\nonumber
&\sum_z [ p(z|x) + p(z|x') ] \sum_{\dot{x}} p(y|\dot{x},z) p(\dot{x}) \leq 1 \text{ and }\\\nonumber
&\sum_z p(z|x) \sum_{\dot{x}} p(y|\dot{x},z) p(\dot{x}) \leq p(x,y)+p(x',y') \text{ and }\\
&\sum_z p(z|x') \sum_{\dot{x}} p(y|\dot{x},z) p(\dot{x}) \leq p(x,y')+p(x',y)
\end{align}
is a necessary condition. Necessary and sufficient conditions for $\epsilon$-bounded immunity on $H$ (i.e., $q(\text{immunity}) \leq \epsilon$) can be obtained much like in Section \ref{sec:cebi}. That is, it suffices to add $\epsilon$ to the right-hand sides of the conditions above and replace $=$ with $\leq$. Finally, we can adapt accordingly the equations in Section \ref{sec:ebbh} to obtain $\epsilon$-bounds on $IB$ and $IH$. Note that the analysis of indirect benefit and harm presented here does not require an RCT, i.e. all the expressions involved can be estimated from just observational data.

\subsection{Example}

This section illustrates the results above with a concrete example borrowed from \citet{Pearl2012}. It concerns the following causal graph:
\begin{center}
\begin{tikzpicture}[inner sep=1mm]
\node at (0,0) (E) {$X$};
\node at (2,0) (D) {$Y$};
\node at (1,0) (Z) {$Z$};
\path[->] (E) edge (Z);
\path[->] (Z) edge (D);
\path[->] (E) edge[bend left] (D);
\end{tikzpicture}
\end{center}
where $X$ represents a drug treatment, $Z$ the presence of a certain enzyme in a patient's blood, and $Y$ recovery. Moreover, we have that
\begin{align*}
p(z|x)=0.75 && p(y|x,z)=0.8\\
&& p(y|x,z')=0.4\\
p(z|x')=0.4 && p(y|x',z)=0.3\\
&& p(y|x',z')=0.2.
\end{align*}
Since $p(x)$ is not given in the original example, we take $p(x)=0.6$.

\citeauthor{Pearl2012} imagines a scenario where the pharmaceutical company plans to develop a cheaper drug that is equal to the existing one except for the lack of direct effect on recovery, i.e. it just stimulates enzyme production as much as the existing drug. Therefore, the probability of benefit of the planned drug is the probability of benefit of the existing drug that is mediated by the enzyme. The company wants to market their drugs by claiming that no one is immune. The sufficient conditions for non-immunity in Equations \ref{eq:sufficient} and \ref{eq:sufficienti} do not hold for the drugs. However, while the existing drug satisfies the necessary condition for non-immunity in Equation \ref{eq:necessary}, the planned drug does not satisfy the corresponding condition in Equation \ref{eq:necessaryi}. Therefore, the company should either abandon their marketing strategy or abandon the plan to develop the new drug and instead focus on trying to confirm non-immunity for the existing drug.

\section{Sensitivity Analysis of Immunity}

In this section, like in the previous section, we assume that the causal graph of the domain under study is available, e.g. from expert knowledge. We also assume that we only have access to observational data, i.e. no RCT is available. Specifically, consider the following causal graph:
\begin{center}
\begin{tikzpicture}[inner sep=1mm]
\node at (0,0) (E) {$X$};
\node at (2,0) (D) {$Y$};
\node at (1,0) (Z) {$Z$};
\path[->] (E) edge (Z);
\path[->] (Z) edge (D);
\path[<->] (E) edge[bend left] (D);
\end{tikzpicture}
\end{center}
which includes potential unmeasured exposure-outcome confounding. Since $p(y_x) = \sum_z p(z|x) \sum_{\dot{x}} E[Y|\dot{x},z] p(\dot{x})$ by the front-door criterion, we can proceed as in the previous section to derive necessary and sufficient conditions for non-immunity. Suppose now that $Z$ is unmeasured or that the effect of $X$ on $Y$ is direct rather than mediated by $Z$. Then, $p(y_x)$ is unidentifiable from observational data \cite{Pearl2009}, and thus we cannot proceed as in the previous section. We therefore take an alternative approach to inform the analyst about the probability of immunity and thereby help her in decision making. In particular, we propose a sensitivity analysis method to bound the probability of immunity as a function of the observed data distribution and some intuitive sensitivity parameters. Our method is an straightforward adaption of the method by \citet{Penna2023}, originally developed to bound the probabilities of benefit and harm.

Let $U$ denote the unmeasured exposure-outcome confounders. For simplicity, we assume that all these confounders are categorical, but our results also hold for ordinal and continuous confounders.\footnote{If $U$ is continuous then sums/maxima/minimima over $u$ should be replaced by integrals/suprema/infima.} For simplicity, we treat $U$ as a categorical random variable whose levels are the Cartesian product of the levels of the elements in the original $U$.

Note that 
\[
p(y_x) = p(y_x | x) p(x) + p(y_x | x') p(x') = p(y | x) p(x) + p(y_x | x') p(x')
\]
where the second equality follows from counterfactual consistency, i.e. $X=x \Rightarrow Y_x = Y$. Moreover,
\[
p(y_x | x') = \sum_u p(y_x | x', u) p(u | x') = \sum_u p(y | x, u) p(u | x') \leq \max_{u} p(y | x, u)
\]
where the second equality follows from $Y_x \ci X | U$ for all $x$, and counterfactual consistency. Likewise,
\[
p(y_x | x') \geq \min_{u} p(y | x, u).   
\]
Now, let us define
\[
M_x = \max_{u} p(y | x, u)
\]
and
\[
m_x = \min_{u} p(y | x, u)
\]
and likewise $M_{x'}$ and $m_{x'}$. Then,
\[
p(x, y) + p(x') m_x \leq p(y_x) \leq p(x, y) + p(x') M_x
\]
and likewise
\[
p(x', y) + p(x) m_{x'} \leq p(y_{x'}) \leq p(x', y) + p(x) M_{x'}.
\]
These equations together with Equation \ref{eq:pi} give
\begin{equation}\label{eq:pisa}
\max \left\{
\begin{array}{cc}
0,\\
p(x') m_x + p(x) m_{x'} - p(y'),\\
p(x') m_x - p(x',y'),\\
p(x) m_{x'} - p(x,y')
\end{array}
\right\}
\leq p(\text{immunity}) \leq
\min \left\{
\begin{array}{cc}
p(x, y) + p(x') M_x,\\
p(x', y) + p(x) M_{x'},\\
p(x') M_x + p(x) M_{x'},\\
p(y)
\end{array}
\right\}
\end{equation}
where $m_x$, $M_x$, $m_{x'}$ and $M_{x'}$ are sensitivity parameters. The possible regions for $m_x$ and $M_x$ are
\begin{equation}\label{eq:pr}
0 \leq m_x \leq p(y|x) \leq M_x \leq 1
\end{equation}
and likewise for $m_{x'}$ and $M_{x'}$.

Our lower bound in Equation \ref{eq:pisa} is informative if and only if\footnote{Note that the second row in the maximum equals the third plus the fourth rows.}
\[
0 < p(x') m_x - p(x',y')
\]
or
\[
0 < p(x) m_{x'} - p(x,y').
\]
Then, the informative regions for $m_x$ and $m_{x'}$ are
\[
p(y'|x') < m_x \leq p(y|x)
\]
and
\[
p(y'|x) \leq m_{x'} < p(y|x').
\]
On the other hand, our upper bound in Equation \ref{eq:pisa} is informative\footnote{Note that we already know that $p(\text{immunity}) \leq p(y)$ by Equation \ref{eq:pi}.} if and only if\footnote{Note that the third row in the minimum equals the first plus the second minus the fourth rows.}
\[
p(x, y) + p(x') M_x < p(y)
\]
or
\[
p(x', y) + p(x) M_{x'} < p(y)
\]
which occurs if and only if $p(y|x) < p(y|x')$ or $p(y|x') < p(y|x)$.\footnote{To see it, rewrite $p(y)=p(x,y)+p(x',y)$ and recall Equation \ref{eq:pr}.} Therefore, our upper bound is always informative, and thus the informative regions for $M_{x}$ and $M_{x'}$ coincide with their possible regions.

\begin{figure}[t]
\centering
\includegraphics[scale=.7]{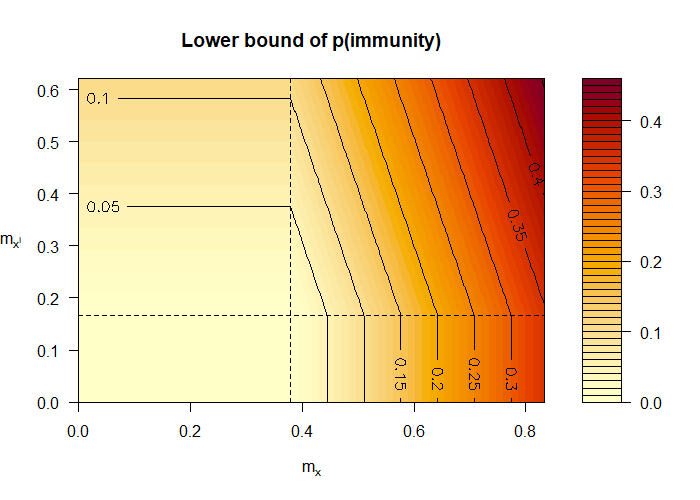}
\includegraphics[scale=.7]{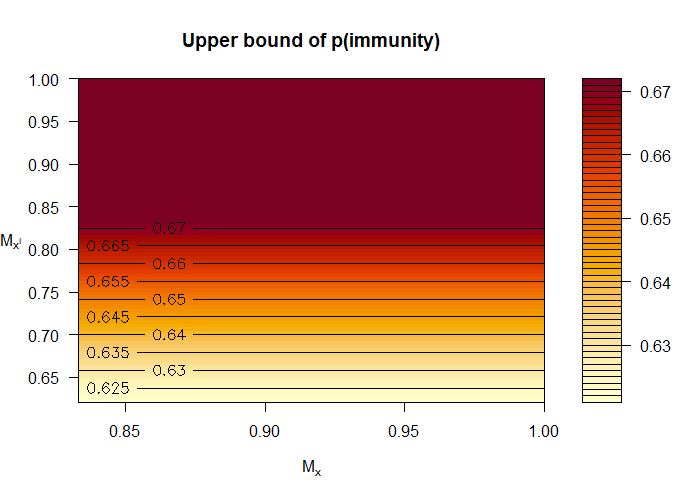}
\caption{Lower and upper bounds of $p(\text{immunity})$ in the example in Section \ref{ex:sa} as functions of the sensitivity parameters $m_x$, $m_{x'}$, $M_x$ and $M_{x'}$.}\label{fig:bounds}
\end{figure}

\subsection{Example}\label{ex:sa}

We illustrate our method for sensitivity analysis of $p(\text{immunity})$ with the following fictitious epidemiological example. Consider a population consisting of a majority and a minority group. Let the binary random variable $U$ represent the group an individual belongs to. Let $X$ represent whether the individual gets treated or not for a certain disease. Let $Y$ represent whether the individual survives the disease. Assume that the scientific community agrees that $U$ is a confounder for $X$ and $Y$. Assume also that it is illegal to store the values of $U$, to avoid discrimination complaints. In other words, the identity of the confounder is known but its values are not. More specifically, consider the following unknown data generation model:
\begin{align*}
p(u)=0.2 && p(x|u)=0.4 && p(y|x,u)=0.9\\
&&&& p(y|x,u')=0.8\\
&& p(x|u')=0.2 && p(y|x',u)=0.2\\
&&&& p(y|x',u')=0.7.
\end{align*}
Since this model does not specify the functional forms of the causal mechanisms, we cannot compute the true $p(\text{immunity})$ \cite{Pearl2009}. However, we can bound it by Equation \ref{eq:pi} and the fact that $p(y_x) = \sum_u p(y | x, u) p(u)$ \cite{Pearl2009}, which yields $p(\text{immunity}) \in [0.42, 0.6]$. Note that these bounds cannot be computed in practice because $U$ is unmeasured.

Figure \ref{fig:bounds} (top) shows the lower bound of $p(\text{immunity})$ in Equation \ref{eq:pisa} as a function of the sensitivity parameters $m_x$ and $m_{x'}$. The axes span the possible regions of the parameters. The dashed lines indicate the informative regions of the parameters. Specifically, the bottom left quadrant corresponds to the non-informative region, i.e. the lower bound is zero. In the data generation model considered, $m_x=0.8$ and $m_{x'}=0.2$. These values are unknown to the epidemiologist, because $U$ is unobserved. However, the figure reveals that the epidemiologist only needs to have some rough idea of these values to confidently conclude that $p(\text{immunity})$ is lower bounded by 0.2. Figure \ref{fig:bounds} (bottom) shows our upper bound of $p(\text{immunity})$ in Equation \ref{eq:pisa} as a function of the sensitivity parameters $M_{x}$ and $M_{x'}$. Likewise, having some rough idea of the unknown values $M_{x}=0.9$ and $M_{x'}=0.7$ enables the epidemiologist to confidently conclude that the $p(\text{immunity})$ is upper bounded by 0.65. Applying Equation \ref{eq:pi} with just observational data produces looser bounds, namely 0 and 0.67. Recall that $p(\text{immunity}) \in [0.42, 0.6]$ in truth.

\section{Discussion}

The analysis in this work can be repeated for $p(\text{doom})$ instead of $p(\text{immunity})$ by simply swapping $y$ and $y'$, and $p(\text{benefit})$ and $p(\text{harm})$. Additionally, the analysis of indirect benefit can be repeated for $p(\text{harm})$ instead of $p(\text{immunity})$ due to Equation \ref{eq:MandP}, and thereby extend the analysis by \citet{MuellerandPearl2023}.

\bibliographystyle{unsrtnat}
\bibliography{sensitivityAnalysis}

\end{document}